\documentclass[prd,amsfonts,twocolumn,nofootinbib,showpacs]{revtex4}
\RequirePackage{graphicx}
\usepackage{enumerate}
\usepackage{bm}
\usepackage{braket}
\usepackage{color}
\usepackage{comment}
\begin{document}
\title{Asymmetric preheating}

\author{Seishi Enomoto}
\affiliation{Department of Physics, University of Florida, Gainesville, Florida 32611, USA}
\author{Tomohiro Matsuda}
\affiliation{Laboratory of Physics, Saitama Institute of Technology,
Fukaya, Saitama 369-0293, Japan}

\begin{abstract}
We study the generation of the matter-antimatter asymmetry
 during bosonic 
preheating, focusing on the sources of the asymmetry.
If the asymmetry appears in the multiplication factor of the resonant
 particle production, the matter-antimatter ratio will grow during preheating.
On the other hand, if the asymmetry does not grow during preheating, one
 has to find out another reason.
We consider several scenarios for the asymmetric preheating to
 distinguish the sources of the asymmetry.
We also discuss a new baryogenesis scenario, in which the asymmetry is
 generated without introducing neither loop corrections nor rotation of
 a field.
\end{abstract}

\maketitle
\section{Introduction}
Violation of CP symmetry is important in particle
physics and cosmology.
The idea of CP violation is crucial in the attempts to explain the
dominance of matter over antimatter in the present Universe and is also
important in the study of the weak interactions of the 
Standard Model(SM).  
If cosmological inflation explains the origin of the large-scale
structure of the cosmos, the temperature after inflation 
is very low.
Therefore, when inflation ends, the Universe must be thermalized by
the decay of the potential energy of the inflaton field.
Although this process is still not well understood, it is believed that
reheating takes place through a process called
``preheating''\cite{Kofman:1997yn}. 
Therefore, asymmetric particle production that works during preheating
is important in cosmology.
The first work on inflationary baryogenesis is given in
Ref.\cite{ClassicBaryo1}, and  the first non-perturbative consideration
of the universe heating after inflation is given in
Ref.\cite{ClassicBaryo2}\footnote{See also Appendix \ref{appendix-history} 
for previous works.}. 
Some secondary effects (e.g, later decay
of a heavy particle\cite{Decay-B} or a false vacuum\cite{Decay-F}) have
been considered by many authors\cite{Allahverdi:2010xz}.
In this paper, we analyze conditions for the asymmetric particle
production in bosonic preheating.
Since preheating uses the Bogoliubov transformation, what is needed
in this work is the asymmetric form of the Bogoliubov transformation and the
evolution equations when CP is violated.
Initially, we have suspected that the difference appears in the multiplication
parameter of the resonant particle production.\footnote{
Because of the
Pauli exclusion principle, fermions do not have such property.}
This is true for the case with an asymmetric initial condition, but in
other cases the source of the asymmetry is different.
We consider effective chemical potential\cite{Cohen:1987vi,Cohen:1988kt},
violation of CP in an initial condition, and CP violation in a kaon-like
system, to understand the origin of the asymmetry.
Then we show that in a simple multi-field model,
kinetic terms and a time-dependent transformation matrix can introduce
the asymmetry.
Using the asymmetric preheating scenario, we show a simple baryogenesis
scenario in Appendix \ref{appendix-baryogenesis}.

\section{Review of bosonic preheating}
First, we briefly explain the traditional bosonic
preheating scenario\cite{Kofman:1997yn} and fix our notations.
We start with the action given by
\begin{eqnarray}
S_0&=&\int d^4 x\sqrt{-g}\left[\partial_\mu\phi^*\partial^{\mu}\phi
-m^2 |\phi|^2+\xi R|\phi|^2 \right].
\end{eqnarray}
Using conformal time $\eta$, we write the metric
$g_{\mu\nu}=a^2(\eta){\rm diag}(1,-1,-1,-1)$ and $R=-6\ddot{a}/a^3$,
where $a$ is the cosmological scale factor and the dot denotes
time-derivative with respect to the conformal time.
It is convenient to define a new field $\chi\equiv a\phi$ and
rewrite the action
\begin{eqnarray}
S_0&=&\int d^4 x \left[|\dot{\chi}|^2
-\omega^2 |\chi|^2\right],
\end{eqnarray}
where
\begin{eqnarray}
\omega^2&\equiv& a^2m^2 +
\left(-\Delta + \frac{\ddot{a}}{a}(6\xi-1)\right).
\end{eqnarray}
Here $\Delta$ is the Laplacian.

We perform decompositions using annihilation ($a,b$) and creation
($a^\dagger,b^\dagger$) operators of ``particle'' and ``antiparticle'';
\begin{eqnarray}
\chi&=& \int \frac{d^3 k}{(2\pi)^{3/2}}\left[
h(\eta) a(\bm{k}) e^{i \bm{k}\cdot \bm{x}}
+g^*(\eta) b^\dagger(\bm{k}) e^{-i \bm{k}\cdot \bm{x}}\right].
\end{eqnarray}
Here $a(\bm{k})$ is the annihilation operator of the positive energy
state.
$b^\dagger(\bm{k})$ is the creation operator of the antiparticle.

For our calculation, we introduce conjugate momenta
$\Pi^\dagger\equiv\dot{\chi}$,
which can be decomposed as
\begin{eqnarray}
\Pi^\dagger&=& \int \frac{d^3 k}{(2\pi)^{3/2}}\left[
\tilde{h}(\eta) a(\bm{k}) e^{i \bm{k}\cdot \bm{x}}
+\tilde{g}^*(\eta) b^\dagger(\bm{k}) e^{-i \bm{k}\cdot \bm{x}}\right].
\end{eqnarray}
To quantize the system, we impose relations
\begin{eqnarray}
\left[\chi(x), \Pi(y)\right]&=&i\delta^3(x-y),
\end{eqnarray}
and
\begin{eqnarray}
\left[a(\bm{k}), a^\dagger(\bm{p})\right]&=&\delta^3(\bm{k}-\bm{p})\\
\left[b(\bm{k}), b^\dagger(\bm{p})\right]&=&\delta^3(\bm{k}-\bm{p}).
\end{eqnarray}
Using $\Pi$,
the second order equation of motion can be written in the first order
equations~\cite{ZS-original}
\begin{eqnarray}
\label{eq-firstorder-b}
\Pi^\dagger-\dot{\chi}&=&0,\nonumber\\
\dot{\Pi}^\dagger+\omega^2 \chi&=&0.
\end{eqnarray}

Following Ref.\cite{ZS-original}, we expand
$h, \tilde{h}$ and $g, \tilde{g}$ in the following way;
\begin{eqnarray}
h&=&\frac{e^{-i\int^\eta \omega d\eta'}}{\sqrt{2\omega}}A_h
+\frac{e^{i\int^\eta \omega d\eta'}}{\sqrt{2\omega}}B_h,\nonumber\\
\tilde{h}&=&\frac{-i\omega e^{-i\int^\eta \omega d\eta'}}{\sqrt{2\omega}}A_h
+\frac{i\omega e^{i\int^\eta \omega d\eta'}}{\sqrt{2\omega}}B_h,
\end{eqnarray}
and
\begin{eqnarray}
g&=&\frac{e^{-i\int^\eta \omega d\eta'}}{\sqrt{2\omega}}A_g
+\frac{e^{i\int^\eta \omega d\eta'}}{\sqrt{2\omega}}B_g,\nonumber\\
\tilde{g}&=&\frac{-i\omega e^{-i\int^\eta \omega d\eta'}}{\sqrt{2\omega}}A_g
+\frac{i\omega e^{i\int^\eta \omega d\eta'}}{\sqrt{2\omega}}B_g,
\end{eqnarray}
where $A$ and $B$ are known as the Bogoliubov coefficients.
For further simplification, we introduce $\alpha$ and $\beta$, which are
defined as
\begin{eqnarray}
\alpha_{h,g}&\equiv& e^{-i\int^\eta\omega d\eta'}A_{h,g}\\
\beta_{h,g}&\equiv& e^{i\int^\eta\omega d\eta'}B_{h,g}.
\end{eqnarray}

Using these equations, one can solve $\alpha$ and $\beta$.
Now the equation of motion can be written as
\begin{eqnarray}
\dot{h}-\tilde{h}&=&0\\
\dot{\tilde{h}}+\omega^2 h&=&0,
\end{eqnarray}
which can be solved for $\dot{\alpha}$ and $\dot{\beta}$ as
\begin{eqnarray}
\dot{\alpha}_h&=&-i\omega \alpha_h
 +\frac{\dot{\omega}}{2\omega}\beta_h\nonumber\\
\dot{\beta}_h&=&i\omega \beta_h
 +\frac{\dot{\omega}}{2\omega}\alpha_h.
\end{eqnarray}

To understand particle production, it is useful to calculate
time-derivatives of $|\alpha|^2$ and $|\beta|^2$.
For real $\omega$, one immediately finds 
\begin{eqnarray}
\label{eq-evoA}
 \frac{d}{dt}|\alpha|^2&=&
  \dot{\alpha}\alpha^*+\alpha\dot{\alpha}^*\nonumber\\
&=& \left(-i\omega \alpha
 +\frac{\dot{\omega}}{2\omega}\beta\right)\alpha^*
+\alpha\left(i\omega \alpha^*
 +\frac{\dot{\omega}}{2\omega}\beta^*\right)\nonumber\\
&=&\frac{\dot{\omega}}{2\omega}\left(\alpha\beta^*+\alpha^*\beta\right)\nonumber\\ 
&=&\frac{\dot{\omega}}{\omega}|\alpha||\beta|\cos{(\theta_\alpha-\theta_\beta)},
\end{eqnarray}
where we defined the phase parameters as $\alpha\equiv
|\alpha|e^{i\theta_\alpha}$.
Here the subscripts $h$ and $g$ are omitted.
The same calculation shows 
\begin{eqnarray}
\label{eq-evoB}
 \frac{d}{dt}|\beta|^{2}&=&
\frac{\dot{\omega}}{\omega}|\alpha||\beta|
\cos{(\theta_\alpha-\theta_\beta)}.
\end{eqnarray}
Comparing these equations, one can see that the evolution equations of
$|\alpha|^2$ and $|\beta|^2$ are identical.
Therefore $|\alpha|^2-|\beta|^2=1$ does not change during preheating, as
required by the definition of these parameters\cite{ZS-original}.

If particle production is not symmetric, $\frac{d}{dt}|\beta_h|^2$  and 
$\frac{d}{dt}|\beta_g|^2$ must be different.
Since matter and antimatter are sharing the same $\omega(t)$, the only
possible way is to find
$\cos{(\theta_{\alpha_h}-\theta_{\beta_h})}\ne\cos{(\theta_{\alpha_g}-\theta_{\beta_g})}$.
Note that $h$ and $g$ are obeying the same equation of motion {\bf as long as
there is no complex parameter}.
The complex parameter, if exists, could or could not be the source of
the asymmetry.
One has to examine if such a parameter generates the asymmetry during
preheating. 
Here we briefly summarize our results.
The chemical potential introduces the complex parameter, but it does not
generate the asymmetry by itself, although the chemical potential is not
compatible with the CPT symmetry.
There are two cases for the asymmetry.
The asymmetric generation is possible in the presence of another interaction,
which explicitly violates the symmetry during the particle production.
This corresponds to the usual spontaneous baryogenesis scenario.
On the other hand, if the initial condition violates the CP condition
described above, even the chemical potential is not needed for the asymmetry.
In this case, although both the chemical potential and the
symmetry (e.g, B) violation is not needed
``during'' the particle production, they have to appear ``before'' the
particle production, since such initial condition will require such
interaction.

In the next section, we study asymmetric preheating in several cases. 
For the first example, we show that the CP violation in the
initial condition can introduce the asymmetry via
$\cos{(\theta_{\alpha_h}-\theta_{\beta_h})}
\ne\cos{(\theta_{\alpha_g}-\theta_{\beta_g})}$. 
For the second example, we see that a complex parameter appears when the
effective chemical potential is introduced to the model.
The chemical potential can indeed split $\alpha_h,
\beta_h$ and $\alpha_g, \beta_g$, but the absolute values of these
parameters remain identical. 
Therefore, in this model, the chemical potential cannot generate the asymmetry.
For the third example, we consider a kaon-like model. 
At the lowest order of the perturbation, eigenstates
of the equation of motion are the CP eigenstates and also they are the 
symmetric combination of matter and antimatter.
Therefore the asymmetry is not generated at the lowest order.
However, in this model, loop corrections introduce
$\Gamma$, which mixes the symmetric eigenstates and causes the asymmetry
between matter and antimatter.
The last example uses multi-field preheating.
In this model, the original eigenstates are not the CP eigenstates but
still a symmetric combination of matter and antimatter.
Mixing between these eigenstates can start when the transformation
matrix $U$ starts to change.
Although the phases in the mass matrix are not changing in our scenario,
the phases in $U$ can change during preheating since it is calculated as
the function of all the parameters in the mass matrix.
Using numerical calculation, we show that a time-dependent $U$
can cause mixing between the eigenstates and introduce the
matter-antimatter asymmetry.

\section{Asymmetry in preheating}
\subsection{CP violation in the initial condition}
\label{subsec-ini}
From Eq.(\ref{eq-evoA}) and (\ref{eq-evoB}), 
one can read out the evolution equations of 
matter ($\alpha_h,\beta_h$) and antimatter ($\alpha_g,\beta_g$).
Since matter and antimatter are sharing the same $\dot{\omega}/\omega$,
matter and antimatter evolve simultaneously when
$\delta_{h-g}\equiv\cos(\theta_{\alpha_h}-\theta_{\beta_h})
-\cos(\theta_{\alpha_g}-\theta_{\beta_g})=0$.
The relation $\delta_{h-g}=0$ is rigorous when CP is
conserved.
Interestingly, the relation could be violated when CP is not conserved
(temporarily) in the early Universe.\footnote{In the next section we
will see that the chemical potential itself cannot violate the condition.}
Fig.1 shows our result of the numerical calculation.
In this scenario, $n_h= n_g$ is possible in the initial
condition, and no explicit CP-violating interaction is needed in the
effective low-energy action. 
In that sense, the asymmetry production is simply the consequence of the initial
condition $\delta_{h-g}\ne0$, which discriminates the later 
evolution of matter from antimatter.
Although the mechanism is quite different, this scenario reminds us of
Affleck-Dine 
baryogenesis\cite{Affleck-Dine} or oscillons\cite{Lozanov:2014zfa}, in
which baryogenesis is triggered by an initial phase-shift.
Unfortunately, at this moment we have no idea how this condition
is realized in a realistic cosmological scenario.\footnote{
The initial phase of a field can be an arbitrary parameter, but it is
not an explicit CP(C) violation.
The Affleck-Dine baryogenesis uses both the initial phase parameter and
a higher dimensional interaction to start rotation of the field, which
gives an ``asymmetric initial condition''. 
On the other hand, in our case, CP violation in the initial condition
specifically means violation of the relation  
$\delta_{h-g}\equiv\cos(\theta_{\alpha_h}-\theta_{\beta_h})
-\cos(\theta_{\alpha_g}-\theta_{\beta_g})=0$.
Also, to make this phase gap effective, we have to assume $\beta\ne
     0$ at the beginning.
Therefore, the initial condition considered in this paper is different
     from the initial conditions used for the previous works.}
Nevertheless, the idea of violating CP by the initial
condition is fascinating, since preheating always starts with a
non-equilibrium state and the initial condition $\delta_{h-g=0}$ 
will not be valid when CP is not the symmetry of the system before the
beginning of the particle production.
\begin{figure}[t]
\centering
\includegraphics[width=1.0\columnwidth]{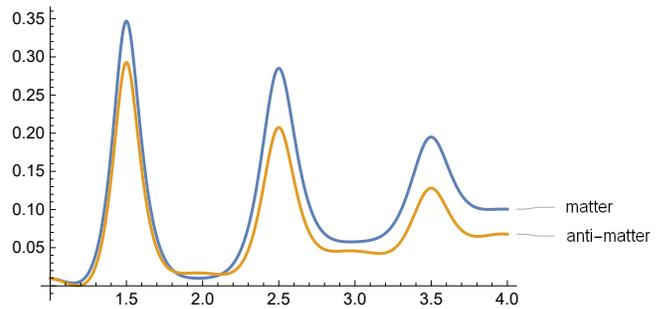}
 \caption{Evolution of number densities are shown.
For the initial condition, we set $|\beta_h|=|\beta_g|=0.1$ but $\delta_{h-g}\ne 0$.
Initially, the number densities are identical but
they start to split.}
\label{fig-phase}
\end{figure}

\subsection{Chemical potential}
\label{sec-chem}
There are a vast variety of interactions that could violate CP.
Among all, the chemical potential would be the simplest.
Using the chemical potential, we can see how preheating works with
such interaction.\footnote{In this
paper, ``chemical potential'' specifically denotes the term given by
Eq.(\ref{chemical-pot-eq}).}

To understand more about the matter-antimatter asymmetry during preheating,
we consider a new scalar field $\varphi$
and its derivative coupling to the current\footnote{We assumed the
charge $q=1$ for $\chi$.}
$J^\mu\equiv -i(\chi \partial^\mu\chi^*-\chi^*\partial^\mu\chi)$,
\begin{eqnarray}
\label{chemical-pot-eq}
{\cal L}_c&=&-\frac{\partial_\mu \varphi}{M_*}J^\mu.
\end{eqnarray}
This term is commonly used in spontaneous
baryogenesis~\cite{Cohen:1987vi, Cohen:1988kt} and applied to fermionic
preheating in 
Ref.\cite{ArmendarizPicon:2007iv, Pearce:2015nga} when the interaction
(or the Majorana mass) other than the chemical potential plays an important
role.
In this paper, we are not arguing the source of this effective chemical
potential.
Also, significant time-dependence of $\mu_\chi$ is not considered, since
in this section we are not arguing the dynamical effect of a 
time-dependent chemical potential.
Assuming a homogeneous background, we have
\begin{eqnarray}
{\cal L}_c&=&-\frac{\dot{\varphi}}{M_*}J^0\\
&\equiv& \mu_\chi (n_\chi-\bar{n}_\chi).
\end{eqnarray}
This term introduces chemical potential
$\mu_\chi\equiv-\dot{\varphi}/M_*$, which is expected to bias the matter
and the antimatter densities.

After adding chemical potential to $S_0$, field equations are changed.
Using
\begin{eqnarray}
\frac{d}{d\eta}\left(\frac{\partial {\cal L}}{\partial
		\dot{\chi}^*}\right)
-\frac{\partial {\cal L}}{\partial \chi^*}&=&0
\end{eqnarray}
for
\begin{eqnarray}
{\cal L}&=&\dot{\chi}\dot{\chi}^* -\omega^2 |\chi|^2
-i\mu_\chi \left(\chi \dot{\chi}^*-\chi^* \dot{\chi}\right),
\end{eqnarray}
we find
\begin{eqnarray}
\label{eq-of-mo-boson}
\ddot{\chi}-2i\mu_\chi\dot{\chi}+(\omega^2-i\dot{\mu}_\chi)\chi&=&0.
\end{eqnarray}
There are two terms which might cause differences.
One is $-2i\mu_\chi\dot{\chi}$, and the other is $-i\dot{\mu}_\chi \chi$.
If one assumes that the chemical potential is not changing with time, one can
simply assume $\dot{\mu}_\chi\simeq 0$.\footnote{This
assumption is made for a single particle production process of
preheating, which has a much shorter timescale compared with
$\dot{\mu}$.
This assumption matches with our focus in this paper.
Introducing off-diagonal interactions $\sim \Lambda \chi^2+h.c$ in
addition to the chemical potential, and taking the opposite limit, 
asymmetry can be produced in terms of the Bogoliubov
transformation\cite{Dolgov:1996qq}.}
Then the equation of motion can be written as
\begin{eqnarray}
\dot{h}-\tilde{h}-i\mu_\chi h&=&0\nonumber\\
\dot{\tilde{h}}+\omega^2h -i\mu_\chi\tilde{h}&=&0,
\end{eqnarray}
where a complex parameter ($\sim i\mu_\chi$) appears.
The imaginary part discriminates the matter equation from the antimatter
equation.

The above equations can be written using $\alpha$ and $\beta$.
One can solve these equations for $\dot{\alpha}$ and $\dot{\beta}$ to
find
\begin{eqnarray}
\dot{\alpha}_h&=&-i(\omega-\mu_\chi)\alpha_h
 +\frac{\dot{\omega}}{2\omega}\beta_h\nonumber\\
\dot{\beta}_h&=&\frac{\dot{\omega}}{2\omega}\alpha_h+i(\omega+\mu_\chi)\beta_h.
\end{eqnarray}
and
\begin{eqnarray}
\dot{\alpha}_g&=&-i(\omega+\mu_\chi)\alpha_g
 +\frac{\dot{\omega}}{2\omega}\beta_g\nonumber\\
\dot{\beta}_g&=&\frac{\dot{\omega}}{2\omega}\alpha_g+i(\omega-\mu_\chi)\beta_g.
\end{eqnarray}

Although the equations are discriminated by the chemical potential, it
is still not clear if the physical 
quantities (e.g, $|\beta|^2$) are different between matter and antimatter.
Indeed, one can calculate the behavior of $|\beta|^2$ to find that the
evolution of $|\beta_h|^2$ and $|\beta_g|^2$ are identical in this
case.
As we have seen in Sec.\ref{subsec-ini}, violation of
$\delta_{h-g}=0$ is crucial for the asymmetric evolution.
Although chemical potential differentiates $\alpha$ and $\beta$,
it does not violate $\delta_{h-g}=0$ in this model.
We also show our numerical calculation in Fig.2, which clearly 
shows that the chemical potential does not violate
$\delta_{h-g}=0$.

Our result suggests that the ``asymmetric preheating'' requires more
complex setups for the CP violation, or an explicit symmetry violation
in the interaction, such as ${\cal L}_{int}\sim \chi^n+h.c.$.
However, reflecting on the current status of CP violation in the SM,
realistic CP violation could not be a simple story.\footnote{The model
with the chemical potential and the explicit CP-violating interaction
(Hermite) gives a simple story, which corresponds to the spontaneous
baryogenesis scenario\cite{Dolgov:1996qq}.} 
Therefore, it is useful to start with a model in which the mechanism of
CP violation is already established and its consequences are widely known.
For this purpose, we choose a K meson (kaon) for our discussion.
\begin{figure}[t]
\centering
\label{fig-chemical}
\includegraphics[width=1.0\columnwidth]{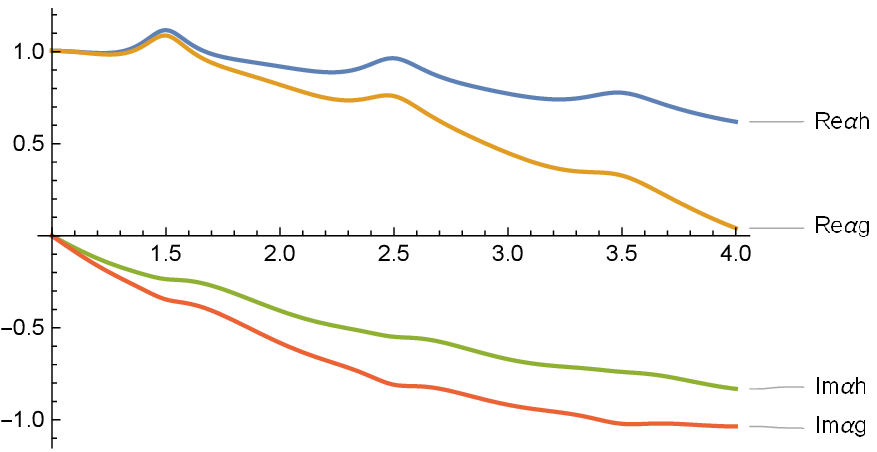}
\includegraphics[width=1.0\columnwidth]{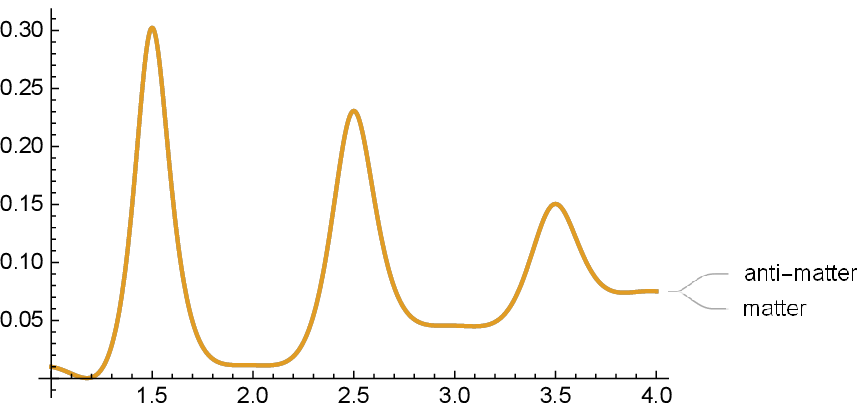}
\includegraphics[width=1.0\columnwidth]{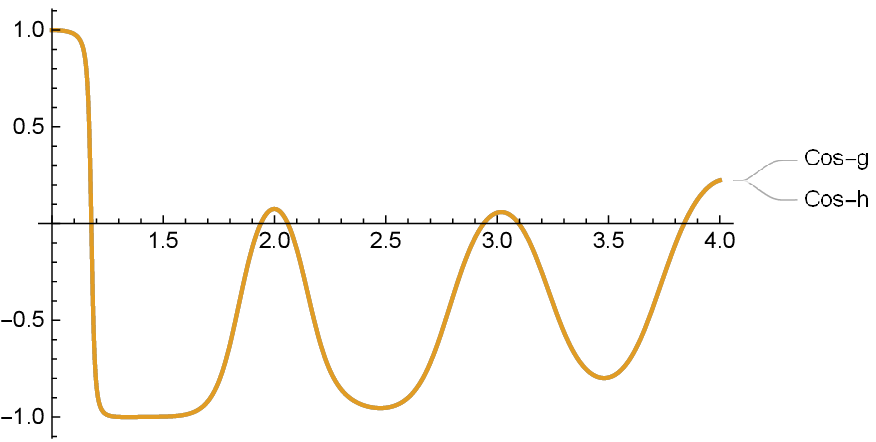}
 \caption{The effect of chemical potential is shown.
The top graph shows that the real and the imaginary parts start to split
 when $\mu\ne 0$. 
The middle shows no difference in $|\beta|$.
The bottom gives $\cos(\theta_{\alpha}-\theta_{\beta})$ for $h$ and $g$,
 which are identical during evolution.}
\end{figure}

\subsection{CP-violating interaction (without the chemical potential)}
Perhaps the most famous CP violation can be found for a kaon,
which denotes a group of mesons distinguished by strangeness.
In the quark model of the SM, they are bound states of a strange quark and an up or
down antiquark.
We consider $K^0=d\bar{s}, \bar{K}^0=\bar{d}s$, which are neutrally
charged and distinguishes matter and antimatter.
Since CP symmetry exchanges matter and antimatter, eigenstates of CP are
$K_1=(\ket{K^0}+\ket{\bar{K}^0})/\sqrt{2}$(CP-even) and
$K_2=(\ket{K^0}-\ket{\bar{K}^0})/\sqrt{2}$(CP-odd).
If CP were conserved, these states could be the eigenstates of weak interaction.
Of course, in reality, they are not the weak eigenstates and therefore CP is
not conserved.
Note that in this model the statement ``CP is conserved during evolution'' is
equivalent to ``CP eigenstates $K_1$ and $K_2$ do not mix during evolution''.
In other words, $K_1$ and $K_2$ can never be the eigenstates of the
equation if the evolution equation violates CP.
This point is crucial for the later argument.

The Schr\"odinger equation for the state
 $\psi_0^T\equiv (K^0,\bar{K}^0)$ can be given by
\begin{eqnarray}
i\frac{d}{dt}\psi_0&=&H\psi_0,\nonumber\\
\left(
\begin{array}{cc}
H_{11} & H_{12}\\
H_{21} & H_{21}
\end{array}
\right)&=& 
\left(
\begin{array}{cc}
M & \Delta_M\\
\Delta_M^* & M
\end{array}
\right).
\end{eqnarray}
However, since a kaon decays during evolution, one has to introduce 
$\Gamma$ as $H\rightarrow H-i\Gamma$, where 
\begin{eqnarray}
\Gamma&=& 
\left(
\begin{array}{cc}
\Gamma_d & \Gamma_\Delta\\
(\Gamma_{\Delta})^*& \Gamma_d
\end{array}
\right)
\end{eqnarray}
CPT transformation ensures $(H-i\Gamma)_{11}=(H-i\Gamma)_{22}$.
Note that $i\Gamma$ is anti-Hermite.

Let us choose the definition in which kaon has real $\Delta_M$.
Without $\Gamma$, the two eigenstates of the equation are
$K_1$ and $K_2$.
If $\Gamma$ is not real, one can decompose it as
$\Gamma_\Delta=\Gamma_\Delta^R+i\Gamma_\Delta^I$,
which gives
\begin{eqnarray}
(H-i\Gamma)_{12}&=&\Delta_M-i(\Gamma_\Delta^R+i\Gamma_\Delta^I)\nonumber\\
(H-i\Gamma)_{21}&=&\Delta_M-i(\Gamma_\Delta^R-i\Gamma_\Delta^I).
\end{eqnarray}
The Schr\"odinger equation for $\hat{\psi}^T\equiv(K_1,K_2)$ becomes
\begin{eqnarray}
i\frac{d}{dt}\hat{\psi}&=&\hat{H}\hat{\psi},\nonumber\\
\hat{H}&=& 
\left(
\begin{array}{cc}
H_{11}-\hat{\Delta} & -\Gamma^I_\Delta\\
-\Gamma^I_\Delta& H_{11}+\hat{\Delta}
\end{array}
\right)\nonumber,
\end{eqnarray}
where $\hat{\Delta}\equiv \Delta_{M} -i\Gamma^R_\Delta$.
The off-diagonal elements $\sim \Gamma_\Delta^I$ measure the CP violation
since it introduces the mixing between the CP eigenstates.

Comparing diagonal elements, one will find a difference proportional to 
$\hat{\Delta}$.
Therefore, the evolution of $K_1$ and $K_2$ are distinguishable.
Although this means that production of $K_1$ and $K_2$ are not
equivalent during preheating, it does not lead to the production of the
asymmetry. 
There are two major problems, which have to be solved.
First, particle production during preheating has to be considered for the eigenstates of the equation.
Therefore, what is produced (directly) during preheating is neither
$K_1$ nor $K_2$.
Second, even if $K_1$ and $K_2$ are produced and their number densities
are different,
both $K_1$ and $K_2$ are one-to-one mixed states of matter ($K_0$) and
antimatter ($\bar{K}_0$).
In this paper, these states are called ``symmetric eigenstates'', because
they do not generate the asymmetry as far as there is no asymmetry in
their decay rates. 

Sometimes a kaon is analyzed using another pair, the two eigenstates of
the above equation, $K_L$ and $K_S$.
They are called weak eigenstates.
Although they do not form orthogonal eigenstates,
preheating must be calculated for the weak eigenstates, since they are
the eigenstates of the equation of motion.

The above argument for a kaon is very useful for our calculation.
To see more details of the model in the light of preheating, 
we first reconsider complex off-diagonal element ($\sim \Delta_M$) and then
introduce complex $\Gamma_\Delta$.

\subsubsection{Complex off-diagonal elements with $\Gamma=0$}
Now consider complex $\Delta(=\Delta_M)$ and find
eigenstates of the equation.
We have
\begin{eqnarray}
\left(
\begin{array}{cc}
H_{11} & H_{12}\\
H_{21} & H_{21}
\end{array}
\right)&=& 
\left(
\begin{array}{cc}
M& \Delta\\
\Delta^* & M
\end{array}
\right).
\end{eqnarray}
The eigenvectors of the matrix are given by
\begin{eqnarray}
\left(-\frac{e^{i\theta_\Delta}}{\sqrt{2}}, \frac{1}{\sqrt{2}}\right),
\left(\frac{e^{i\theta_\Delta}}{\sqrt{2}}, \frac{1}{\sqrt{2}}\right),
\end{eqnarray}
where $\Delta\equiv |\Delta| e^{i\theta_\Delta}$.
Their eigenvalues are
\begin{eqnarray}
M-|\Delta|, M+|\Delta|.
\end{eqnarray}
Of course, the two eigenvectors are orthogonal.
In this case, the eigenstates are symmetric eigenstates.
Obviously, this is due to the CPT transformation, which strictly
decides the diagonal elements of the original matrix to be 
$H_{11}=H_{22}$. 
For the system with only matter and antimatter, 
this dilemma is quite serious.

\subsubsection{Off-diagonal $\Gamma$}
Now we introduce $\Gamma$ to the above model.
Since we are seeing matter-antimatter bias in the
eigenstates, we focus on the off-diagonal elements of $\Gamma$
(and disregard diagonal elements of $\Gamma$).
We have 
\begin{eqnarray}
\left(
\begin{array}{cc}
H_{11} & H_{12}\\
H_{21} & H_{21}
\end{array}
\right)&=& 
\left(
\begin{array}{cc}
M& \Delta_M+i\Gamma_\Delta\\
\Delta_M^* +i\Gamma_\Delta^*& M
\end{array}
\right).
\end{eqnarray}
Eigenvectors are
\begin{eqnarray}
\left(\pm\frac{r}{\sqrt{1+r^2}},\frac{1}{\sqrt{1+r^2}}\right),
\end{eqnarray}
where $r\equiv
\sqrt{\frac{\Delta+i\Gamma_\Delta}{\Delta^*+i\Gamma_\Delta^*}}$.
This parameter is commonly used to measure the CP-violation in a
kaon.

Their eigenvalues are
\begin{eqnarray}
M\pm\sqrt{(\Delta+i\Gamma_\Delta)(\Delta+i\Gamma_\Delta^*)}.
\end{eqnarray}
Now we have eigenstates which are a set of biased mixed states of matter
and antimatter (asymmetric states).

Note that in this model $r=1$ (symmetric eigenstate) is realized when both 
$\Gamma_\Delta$ and $\Delta$ are real, or either $\Gamma_\Delta$ or
$\Delta$ vanishes. On the other hand,
$r\ne 1$(asymmetric eigenstate) can be realized if either $\Gamma_\Delta$ or
$\Delta$ is complex and both $\Gamma_\Delta$ and
$\Delta$ do not vanish.

During preheating, the two eigenstates of the equation are produced.
They are expected to have different number densities.
However, since the matter-to-antimatter ratio ($r$) is identical between
these eigenstates, the ratio between matter and antimatter does not depend on
their number densities.
Although the Kaon is not giving a realistic baryogenesis scenario, 
it can be considered as a typical matter-antimatter system.
For example, a similar discussion can be found in leptogenesis, although
it has been discussed for the decaying process, not for the preheating
(production) process.
The CP violation in the decays of heavy singlet neutrinos, which arises
from the wave function mixing, has been calculated in
Ref.\cite{Covi:1996wh, Flanz:1996fb}.
In their analyses, CP violation in the wave function mixing is found to
generate a similar contribution (bias) to the eigenstates of the heavy singlet
neutrinos\footnote{Since the contribution appears for the ``mixing'',
one has to introduce at least two singlet neutrinos for the model.}.
Although originally the bias caused by the wave function mixing was
discussed for the decay process, the same bias can generate the
asymmetry during the production process of preheating.
Namely, if one of the heavy singlet neutrinos has a field-dependent
mass and is generated during preheating, it causes the bias $r\ne 0$.
To avoid thermalization after preheating, which washes out the 
lepton number and resets the initial condition, one has to assume
that the reheating temperature $T_R$ is low enough to avoid
thermalization of the singlet fermions.

\subsection{Distinguishing sources of the asymmetry}
We have seen that in some cases the eigenstates are biased during preheating.
Since the equation of motion is diagonalized by the eigenstates, 
these particles are produced independently during preheating.
Their number densities could be different because their eigenvalues are
not always the same.
On the other hand, since these eigenstates are sharing the same $r$,
the asymmetry is uniquely determined by $r$.
{\bf This property is very important.}
In a model where the particle number density grows exponentially during
preheating, 
one can see that the $j$-th scattering gives\footnote{For more details
see the original paper\cite{Kofman:1997yn}.}
\begin{eqnarray}
\label{eq-grow}
n_k^{j+1} &=&e^{-\pi \kappa^2}+(1+2e^{-\pi \kappa^2})n_k^j\nonumber\\
&&-2e^{-\frac{\pi}{2} \kappa^2}\sqrt{1+e^{-\pi \kappa^2}}
\sqrt{n_k^j(n_k^j+1)}\sin \theta^j,\nonumber\\
\end{eqnarray}
where $\kappa^2\equiv \frac{k^2}{k_*^2}\equiv \frac{k^2}{gv}$
and $\theta^j$ is the total phase accumulated by the moment $t=t_j$.
Here $k_*$ gives the typical width of the Gaussian function $e^{-\pi
k^2/k_*^2}$.
The first term gives the particle production without resonance.
The second term is the source of resonant amplification of the number density.
The third term may lead to stochastic behavior.
Eq.(\ref{eq-grow}) shows that {\bf a bias in the second term (in the
multiplication factor 1+2$e^{-\pi \kappa^2}$) will give the growing asymmetry.}
On the other hand, if the model has a bias ($r$) in the eigenstates, it 
leads to {\bf the time-independent asymmetry}.
These results are very useful.
Since preheating is a highly non-linear process, it becomes difficult
to understand the origin of the asymmetry when many fields and various
interactions are introduced. 
If the numerical calculation shows that the asymmetry is nearly
constant from the beginning, one can predict that the asymmetry is mostly
controlled by the bias in the eigenstates.
On the other hand, if the asymmetry grows significantly during
preheating, one can predict that the multiplication factor
$\sim e^{-\pi\kappa^2}$ is the source of the asymmetry.
If the asymmetry behaves somewhat randomly, there could be a mismatch in
the phase (see Section \ref{subsec-ini}).
Therefore, the above properties are giving the lead to the
origin of the asymmetry.\footnote{Because of the Pauli 
blocking, the above discussion does not apply to the fermions.
}

\section{Multi-field extension : Asymmetry without loop corrections}
In the previous section, we have seen how the asymmetry appears in the
system of a complex scalar field. 
For our purpose, we are going to introduce a complex parameter by 
introducing another field and interaction. 

In the kaon-like model, the asymmetry appears when {\bf
interaction} generates $\Gamma$.
Because of $\Gamma$, the original CP-eigenstates are mixed to generate
the biased eigenstates.
 
In this section, we try to construct a model of asymmetric preheating
when loop correction is not the source of the asymmetry.
Instead of using interaction, we consider {\bf kinetic terms} for 
the mixing.
Then, if the mixing is accompanied by a phase, asymmetry may
appear in the eigenstates.
Note that we are {\bf not} considering non-minimal kinetic terms for our
story.
Suppose that the interaction is diagonalized using a transformation matrix
$U$.
If one considers conventional (minimal) kinetic terms, the kinetic terms are already
diagonalized for the original fields. 
Normally, $U$ is not time-dependent and thus
nothing will happen: 
kinetic terms are (usually) diagonal for both the original and
the new fields.
However, during preheating, the matrix can be time-dependent.
In that case, the minimal kinetic term can generate
additional (effective) CP-violating interaction and mixing between
the states.

In this section, we show a simple model in which asymmetric
preheating is realized by the time-dependent matrix $U$.

The simplest model can be given by the following Lagrangian;
\begin{eqnarray}
\label{simplest3by3}
{\cal L}&=&|\partial_\mu \phi|^2 -m_\phi^2|\phi|^2\nonumber\\
&& +\frac{1}{2}(\partial_\mu\eta)^2-\frac{1}{2}m_\eta^2
 \eta^2\nonumber\\
&&-\frac{1}{2}(\epsilon\phi^2+h.c.) -(g\phi\eta+h.c.),
\end{eqnarray}
where $\phi$ is a complex scalar and $\eta$ is a real scalar
and $\epsilon$, $g$ are complex coupling
constants, which are dimensional.\footnote{
In this paper, we have assumed that higher interactions are very small
and negligible. 
Preheating with higher interactions is partially analyzed in
Ref.\cite{Enomoto:2013mla, Enomoto:2014hza,Enomoto:2014cna}.}
Note that the complex phases of $\epsilon$ and $g$ are not removed
simultaneously.\footnote{The situation reminds us of
Cabibbo-Kobayashi-Maskawa(CKM) Matrix in the SM. 
Unless $U$ is given by a real orthogonal matrix, a complex phase will
remain.} 
Here, we choose the definition that makes $\epsilon$ to be real.
Furthermore, we assume that $\phi$'s mass $m_\phi^2$ depends on time
in order to take into account the situation of preheating, for example,
$m_\phi^2=m^2+\lambda\varphi^2(t)$ where $m$ is $\phi$'s original mass,
$\lambda$ is a coupling, and $\varphi$ is an oscillating background field.
The equations of motion are given by
\begin{equation}
 \ddot{\Psi}+\Omega^2\Psi=0 \label{eq:eom_phichi}
\end{equation}
where
\begin{equation}
 \Psi \equiv \left( \begin{array}{cc} \phi \\ \phi^\dagger \\ \eta \end{array} \right),
 \quad \Omega^2 \equiv \left(
  \begin{array}{ccc}
   \omega_\phi^2 & \epsilon & g^*\\
   \epsilon & \omega_\phi^2 & g\\
   g & g^* & \omega_\eta^2
  \end{array}
 \right).
\end{equation}
In order to understand the above equations, we first discuss the
eigenvalues and eigenvectors of the matrix $\Omega^2$.
Although the formula becomes cumbersome, since their eigenvalues are
given by the solutions of a cubic equation,
fortunately, it is easy to find that the eigenvectors
are expressed as $\propto(a_i, a_i^*,1)$ for $i=1,2,3$.
Here $a_i$ is a complex number.
This means that the eigenstates are 
$\psi_i\equiv b_i(a_i\phi+a_i^*\phi^*+\eta),$
where $b_i$ is a normalization factor.
Although these ``eigenstates''\footnote{Remember that these are not the
eigenstates of the equation of motion when $U$ is time-dependent.} are
not CP eigenstates (except for the special points in the parameter space), 
they are symmetric in the sense that they are the one-to-one mixing
states of $\phi$ and $\phi^\dagger$.
The point is that since $\dot{\omega}_\phi(t)\ne 0$ is considered,
the above ``eigenstates'' do not diagonalize the equation of motion
during preheating.
Assume that the mass matrix $\Omega^2$ is diagonalized by a unitary matrix $U$
given by
\begin{equation}
 \omega^2 \equiv \left( \begin{array}{ccc} \omega_1^2 & & \\ & \omega_2^2 & \\ & & \omega_3 \end{array} \right)
  = U^\dagger\Omega^2U,
\end{equation}
where we defined\footnote{Phases in this matrix is not identical to the
phase of the complex parameter $g$. They are determined by all the
parameters in $\Omega^2$. As the result, phases can be time-dependent
when $\dot{\omega}_\phi\ne 0$.}
\begin{equation}
 U^\dagger=\left( \begin{array}{ccc} a_1b_1 & a_1^*b_1 & b_1 \\
  a_2b_2 & a_2^*b_2 & b_2 \\ a_3b_3 & a_3^*b_3 & b_3 \end{array} \right) = U^{-1}.
\end{equation}
Then the equations of motion (\ref{eq:eom_phichi}) becomes 
\begin{equation}
 (U^\dagger\Psi)^{\cdot\cdot}+2\gamma(U^\dagger\Psi)^\cdot+(\omega^2+\gamma^2+\dot{\gamma})(U^\dagger\Psi)=0,
\end{equation}
where we defined
\begin{equation}
 \gamma \equiv U^\dagger\dot{U}.\label{eq:uudot}
\end{equation}
which is anti-Hermite.  
Because of $\gamma$, the equation of motion is not diagonalized by the
``symmetric states''.
Here $U$ depends on time as long as
$\dot{\omega}_\phi\neq0.$\footnote{Note that 
$\dot{\omega}_\phi\neq0$ does not always mean $\dot{U}\neq0$. 
A typical example would be $\omega_{\phi}(t)=\omega_{\eta}(t),$
in which $U$'s components %does not depend on $\omega_\phi$.
are represented by only constant parameters $\epsilon, g$, i.e., $\dot{U}=0$.
In that case production of the asymmetry is impossible.}
This fact indicates that the asymmetry may appear when $\gamma\ne 0$,
because it mixes the eigenstates and the CP violation may cause the
interference.
If the speculation is correct, the asymmetry between particles and antiparticles would
be enhanced in the non-adiabatic regime, where time-dependence becomes
significant. 
This corresponds to the area where particle production becomes
significant.

\begin{figure}[t]
 \begin{center}
  \includegraphics[width=1.0\columnwidth]{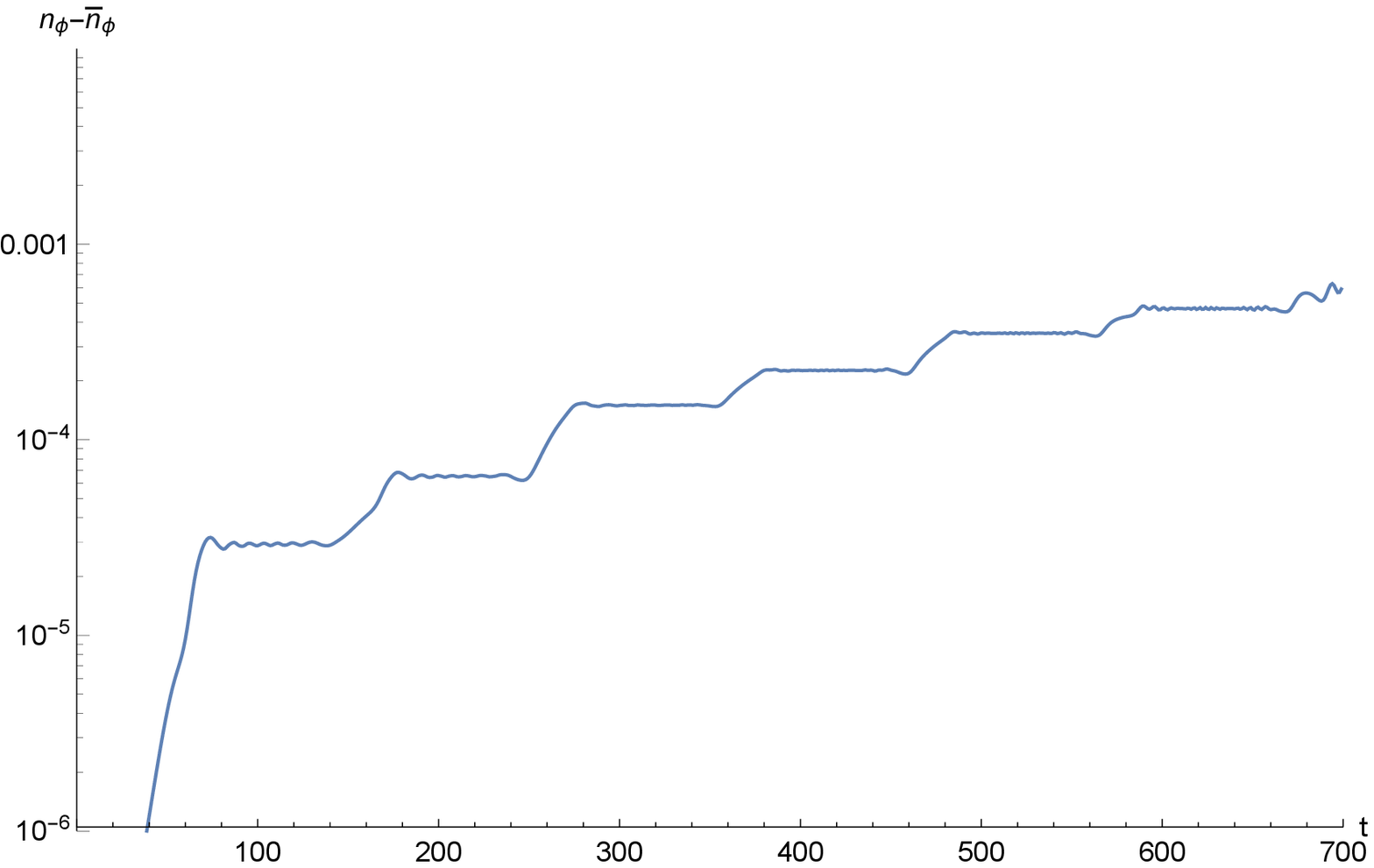}
  \includegraphics[width=1.0\columnwidth]{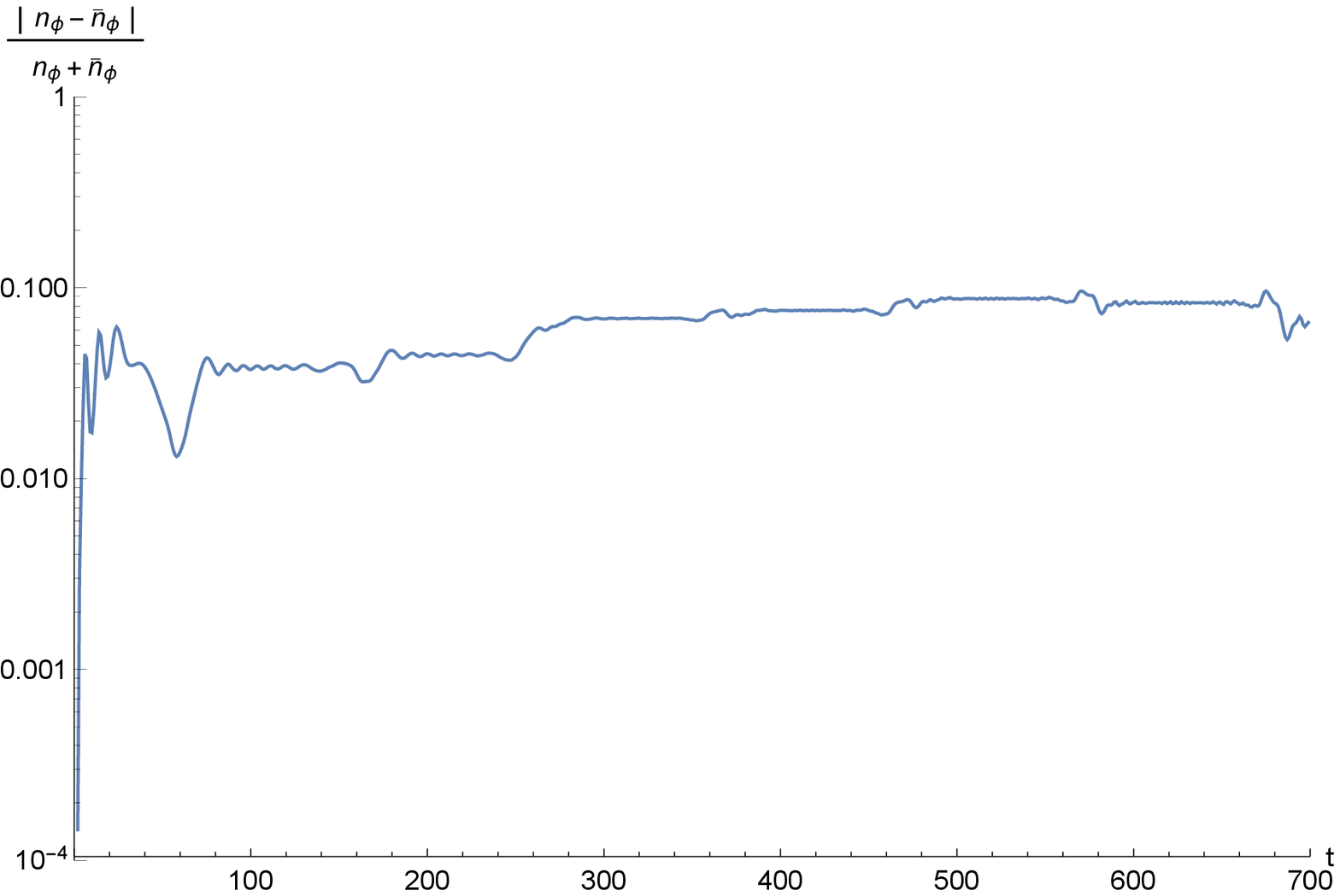}
  \caption{Time evolution of the net number (upper) and the asymmetry
  (bottom) of $\phi$.
   We assumed $m_\phi^2(t)=m^2+\lambda\varphi^2(t)$ and chose parameters as
  $m=0.15, \lambda=4, \varphi(t)=\cos0.03t,
  m_\eta=0.1, \epsilon=10^{-4}, g=10^{-2}i$. From the second plot, we
  can see that the ratio behaves like a constant from the
  beginning. This clearly shows that the bias in the eigenstates is
  controlling the asymmetry.}
  \label{fig:asymmetry}
  \end{center}
\end{figure}

Since the asymmetry is evaluated by
\begin{equation}
 n_\phi-\bar{n}_\phi = \frac{1}{V}\int d^3x \:
  i(\langle \phi^\dagger \dot{\phi} \rangle - \langle \phi \dot{\phi}^\dagger \rangle),
\end{equation}
where $V$ is a volume of the system,
in principle one can follow the evolution of the asymmetry by solving eq.(\ref{eq:eom_phichi}).
The problem is that it is not an easy task to obtain an analytic
solution of eq.(\ref{eq:eom_phichi}).
Instead, we show the results of our numerical calculation in Fig.\ref{fig:asymmetry},
which clearly shows that the asymmetry is generated but the ratio does
not grow when the total number density grows.
This result reminds us of the kaon-like model, in which the ratio is
expected to be constant during preheating.

Finally, we emphasize that the asymmetry has been generated by
the simple oscillating background, not by a rotation.
This is possible because the parameters in $U$ is the complex
functions of the original parameters in $\Omega$.

\section{Conclusions and discussions}
In this paper, we studied the possibility of generating asymmetry between
matter and antimatter 
with the CP violating models due to the parametric resonance.
Effective chemical potential, CP violation in the initial condition, CP
violation in a kaon-like system (loop corrections) and a time-dependent
transformation matrix are examined.

Violation of the CP initial condition is fascinating since inflation
may start with a non-equilibrium state.
However, at this moment appearance of such initial condition is an optimistic 
option.

If the Hamiltonian is Hermite, the unremovable phases and the
time-dependent transformation matrix play the crucial role.
We considered a multi-field model to show how this idea works.
In this model, the eigenstates are not the CP eigenstates, but {\bf without
$\dot{U}\ne 0$, eigenstates of the equation 
of motion are still symmetric.} 
Therefore, the matter-antimatter symmetry still remains when $U$ is constant. 
To achieve $\dot{U}\ne 0$, we considered $\dot{\omega}_\phi\ne 0$.
The mass of the additional scalar field does not have to
be time-dependent but must not be identical to the complex field mass,
since in that case, $U$ does not depend on time.
In this story, no special structure is needed for the kinetic term.
In this scenario, the matter-antimatter symmetry appears not from the
interaction but from the kinetic term.

In the latter half of this paper, we have tried to reveal the origin of
the asymmetry focusing on the bias in the eigenstates.
Such bias may appear either from the loop 
corrections or from the dynamical effects.
We have seen that in the former case the interference between
corrections gives $\Gamma\ne 0$, which is crucial for the asymmetry.
In the latter case, $\gamma\ne 0$ is crucial for the asymmetry.

\section{Acknowledgments}
SE is supported by the Heising-Simons Foundation grant No 2015-109.

\appendix
\section{Baryogenesis}
\label{appendix-baryogenesis}
In this appendix, we are going to discuss how baryogenesis
can be realized using $\gamma\ne 0$.
For simplicity, it would be useful to follow the scenario considered in
Ref.\cite{Decay-B}.
The original setup, simple chaotic inflation, could not be explaining
the current cosmological parameters, but the inflaton oscillation after
inflation is still generic.
Here, the significant difference from the traditional scenario is 
in the source of the asymmetry.
In the traditional scenario of Grand Unified Theories (GUT)
baryogenesis, no asymmetry is assumed for the initial number densities
of the heavy fields.
Instead, the asymmetry is generated by the asymmetric decay width, which
is due to the loop corrections.
In our scenario of asymmetric preheating, the number densities of the
heavy fields are not symmetric, while no asymmetry is assumed
 for the decay rates.

We start our discussion with the traditional Sakharov
conditions\cite{Sakharov}: B violation, out of thermal equilibrium and
C, CP violation.

\subsection{B violation}
In the GUT, there are heavy gauge bosons $X^\mu$
and Heavy Higgs bosons $Y$ with couplings to quarks and leptons.
The couplings are schematically written as $X qq, X\bar{q}\bar{l}$ (and
similar for Y), which implies that baryon number cannot be assigned
consistently to these heavy bosons.
In supersymmetric models, one could avoid R-parity and introduce
dimension-four baryon number violating operators, such as
\begin{eqnarray}
\tilde{q}_t^a q_b^b q_s^c\epsilon_{abc},
\end{eqnarray}
where $\tilde{q}_t$ is the bosonic partner of the top quark (squark).
This interaction could exist without violating the constraint from the
proton decay if lepton number was independently conserved for some
reason
(i.e, $p\rightarrow \pi^+ \nu$ and $p\rightarrow \pi^0 e^+$ are forbidden by
the lepton number conservation.)
Another way of realizing a similar scenario is to consider the asymmetric
generation of a singlet sneutrino.

Traditional GUT baryogenesis requires multiple decay process
since without such process the numbers of $q$ and $\bar{q}$ could eventually
indistinguishable even if $\Gamma(X\rightarrow qq)\ne
\Gamma(\bar{X}\rightarrow \bar{q}\bar{q})$.
Moreover, to realize the asymmetry in the decay rate, one has to
introduce multiple fields, which are responsible for the interference.
Although these properties could be ``natural'' in the GUT, they are making
the scenario rather complicated.

The simplest way to realize baryogenesis in our scenario is to assume that
one of these heavy fields ($X,Y$ or $\tilde{q}_t$) are generated by the
asymmetric preheating.
Then, $n_X \ne n_{\bar{X}}$ can be realized for the heavy field.

Unlike the traditional scenario, small decay width
 is not excluding our scenario, since the source of the
asymmetry is not the interference caused by the loop corrections.

As in Ref.\cite{Dolgov:1994zq}, we consider a simple model involving a
complex scalar field $\phi$ and fermion fields $Q$ and $L$ with the
Lagrangian density given by
\begin{eqnarray}
{\cal L}&=& {\cal L}_{ap}+i\bar{Q}\gamma^\mu\partial_\mu Q+
i\bar{L}\gamma^\mu\partial_\mu L -m_Q \bar{Q}Q-m_L \bar{L}L\nonumber\\
&&+\left(g\phi \bar{Q}L+h.c.\right).
\end{eqnarray}
Note that, in this simple model, $Q$ and $L$ are not the quarks and leptons
in the SM.
They represent heavy fermions, which have other interactions with the
standard model.
We assume that $Q$ and $\phi$ are carrying baryon number $+1$ and $-1$,
respectively.
The baryon number is explicitly violated in ${\cal L}_{ap}$, which is
the same as Eq.(\ref{simplest3by3}),
\begin{eqnarray}
{\cal L}_{ap}&=&|\partial_\mu \varphi|^2+|\partial_\mu \phi|^2 
-m_\phi^2|\phi|^2-m_\varphi^2|\varphi|^2\nonumber\\
&& +\frac{1}{2}(\partial_\mu\eta)^2-\frac{1}{2}m_\eta^2
 \eta^2\nonumber\\
&&-\frac{1}{2}(\epsilon\phi^2+h.c.) -(g\phi\eta+h.c).
\end{eqnarray}
Here, $\varphi(t)$ starts to oscillate after inflation and 
the asymmetric particle production is realized by the time-dependent mass
$m_\phi^2=m_0^2+\lambda \varphi^2$.
One can assume that the potential of $\varphi$ is effectively quadratic
during oscillation. 
Since the heavy fields $\phi$ and $\eta$ are violating the baryon
number, these fields must not be in thermal equilibrium.
Therefore, we assume for the reheating temperature $T_R \ll m_0
<m_\eta$, which does not constitute an additional limiting
factor.\footnote{A similar condition has been used in
Ref.\cite{Decay-B}.}

\subsection{Out of thermal equilibrium}
If a baryon number violating process is still in thermal equilibrium,
the inverse process destroys baryon as fast as it is created.
If the heavy fields are generated in a thermal plasma, this condition is
very important.
On the other hand, if the heavy fields are generated during preheating,
the Universe is already far away from thermal equilibrium.
Therefore, as far as the baryon number violating process does not come
into thermal equilibrium after preheating, the washout process is not
important.
Of course, the sphaleron process is important when it is activated in the
thermal plasma. However, it cannot wash out $B-L$, where $B$ and $L$ are
the baryon and the lepton numbers of the Universe, respectively.
This condition is rather trivial in our scenario.

\subsection{C,CP violation}
Let us consider the Lagrangian given by Eq.(\ref{simplest3by3}).
CP is violated when both $\epsilon$ and $g$ are complex.
These CP phases can be changed by using the phase rotation of $\phi$.
After rotation, one can find the Lagrangian written with real $\epsilon$
and complex $g$.
This is the starting point of our baryogenesis scenario.
Obviously, the asymmetry disappears when all these parameters become real
after the phase rotation.
 
In our simple scenario, the complex scalar field $\phi$ is the heavy
field that is responsible for baryogenesis.
One might think that tuning parameters, it is possible to make the real field 
a candidate of the dark matter.
Such scenario could be fascinating, but in this paper, we are simply
assuming that the real field is heavier than the complex scalar field and
it decays fast.
Then, since the real field does not generate asymmetry, its decay
dilutes the baryon number of the Universe.
However, since our scenario can expect asymmetry 
$\epsilon_\phi\equiv\frac{n_\phi-n_{\bar{\phi}}}{n_\phi+n_{\bar{\phi}}}$
up to $O(0.1)$ for the maximum CP violation, and the 
initial density of the real field can be the same order as the heavy field, 
it is easy to generate the required baryon number of the Universe using
the mechanism.

\section{Multi-field preheating and the asymmetry}
\label{appendix-history}
A scenario for multi-field bosonic preheating has been discussed
by Funakubo et.al. in Ref.\cite{Funakubo:2000us}.
The model discussed in this paper partially overlaps with our study.
However, they concluded that at least two complex fields
and time-dependence of ``not only mass terms but also phases in the
mass matrix parameters'' are needed in order to generate charge asymmetry.
In the light of cosmology, these conditions are very stringent.

Our discussion is different from Ref.\cite{Funakubo:2000us}.
We noticed that the complexity of preheating in a multi-field system makes
the calculation quite unclear and it disturbs finding the required
condition for the asymmetry.
To make the condition for the asymmetry clear in multi-field preheating,
we have used the eigenstates of the equation to find that the dynamical mixing
accompanied by a complex phase can lead to the bias between matter and
antimatter. 
To realize this scenario, one has to introduce at least one complex
and one real scalar fields, which couple via complex interaction.
Also, the complex scalar should have CP-violating interaction $\sim
\lambda_n \phi^n+h.c$.
With these simple setups, we have shown that asymmetry can be generated by a
time-dependent mass of the complex scalar field.
Our model does not need any time-dependent parameter other than the
scalar mass, which can make the scenario of asymmetric preheating
very simple.
The trick is in the time-dependent matrix $U$.
Since the phases in $U$ is determined by all the parameters in the mass
matrix, not only the phase of $g$ but also $m_\phi$ can change the
phases.
As the result, one can introduce effective phase rotation in $U$ without
introducing the phase rotation of the original parameters in the mass matrix.


\begin{thebibliography}{1}
\bibitem{Kofman:1997yn}
  J.~H.~Traschen and R.~H.~Brandenberger,
  ``Particle Production During Out-of-equilibrium Phase Transitions,''
  Phys.\ Rev.\ D {\bf 42}, 2491 (1990);
  L.~Kofman, A.~D.~Linde and A.~A.~Starobinsky,
  ``Reheating after inflation,''
  Phys.\ Rev.\ Lett.\  {\bf 73}, 3195 (1994)
  [hep-th/9405187];
  L.~Kofman, A.~D.~Linde and A.~A.~Starobinsky,
  ``Towards the theory of reheating after inflation,''
  Phys.\ Rev.\ D {\bf 56}, 3258 (1997)
  [hep-ph/9704452].
\bibitem{ClassicBaryo1}
  A.~D.~Dolgov and A.~D.~Linde,
  ``Baryon Asymmetry in Inflationary Universe,''
  Phys.\ Lett.\  {\bf 116B} (1982) 329;
\bibitem{ClassicBaryo2}
 A.~D.~Dolgov and D.~P.~Kirilova,
  ``On Particle Creation By A Time Dependent Scalar Field,''
  Sov.\ J.\ Nucl.\ Phys.\  {\bf 51} (1990) 172
   [Yad.\ Fiz.\  {\bf 51} (1990) 273].
\bibitem{Decay-B}
 E.~W.~Kolb, A.~D.~Linde and A.~Riotto,
  ``GUT baryogenesis after preheating,''
  Phys.\ Rev.\ Lett.\  {\bf 77} (1996) 4290,
  [hep-ph/9606260].
\bibitem{Decay-F}
 G.~W.~Anderson, A.~D.~Linde and A.~Riotto,
  ``Preheating, supersymmetry breaking and baryogenesis,''
  Phys.\ Rev.\ Lett.\  {\bf 77}, 3716 (1996),
  [hep-ph/9606416];
  J.~Garcia-Bellido, D.~Y.~Grigoriev, A.~Kusenko and M.~E.~Shaposhnikov,
  ``Nonequilibrium electroweak baryogenesis from preheating after inflation,''
  Phys.\ Rev.\ D {\bf 60}, 123504 (1999)
  [hep-ph/9902449].
\bibitem{Allahverdi:2010xz} 
  R.~Allahverdi, R.~Brandenberger, F.~Y.~Cyr-Racine and A.~Mazumdar,
  ``Reheating in Inflationary Cosmology: Theory and Applications,''
  Ann.\ Rev.\ Nucl.\ Part.\ Sci.\  {\bf 60}, 27 (2010)
  [arXiv:1001.2600 [hep-th]].
\bibitem{Cohen:1987vi} 
  A.~G.~Cohen and D.~B.~Kaplan,
  ``Thermodynamic Generation of the Baryon Asymmetry,''
  Phys.\ Lett.\ B {\bf 199}, 251 (1987).
\bibitem{Cohen:1988kt}
  A.~G.~Cohen and D.~B.~Kaplan,
  ``Spontaneous Baryogenesis,''
  Nucl.\ Phys.\ B {\bf 308}, 913 (1988).
\bibitem{ArmendarizPicon:2007iv}
  C.~Armendariz-Picon, M.~Trodden and E.~J.~West,
  ``Preheating in derivatively-coupled inflation models,''
  JCAP {\bf 0804} (2008) 036
\bibitem{Pearce:2015nga}
  L.~Pearce, L.~Yang, A.~Kusenko and M.~Peloso,
  ``Leptogenesis via neutrino production during Higgs condensate relaxation,''
  Phys.\ Rev.\ D {\bf 92} (2015) no.2,  023509

\bibitem{ZS-original}
Y. B. Zeldovich and A. A. Starobinsky, Particle production and vacuum
	polarization in an anisotropic gravitational field,
	Sov. Phys. JETP 34 (1972) 1159.
\bibitem{Affleck-Dine}
  I.~Affleck and M.~Dine,
  ``A New Mechanism for Baryogenesis,''
  Nucl.\ Phys.\ B {\bf 249}, 361 (1985).
\bibitem{Lozanov:2014zfa}
  K.~D.~Lozanov and M.~A.~Amin,
  ``End of inflation, oscillons, and matter-antimatter asymmetry,''
  Phys.\ Rev.\ D {\bf 90} (2014) no.8,  083528
  [arXiv:1408.1811 [hep-ph]].
\bibitem{Dolgov:1996qq}
  A.~Dolgov, K.~Freese, R.~Rangarajan and M.~Srednicki,
  ``Baryogenesis during reheating in natural inflation and comments on spontaneous baryogenesis,''
  Phys.\ Rev.\ D {\bf 56} (1997) 6155
  [hep-ph/9610405].
\bibitem{Covi:1996wh}
  L.~Covi, E.~Roulet and F.~Vissani,
  ``CP violating decays in leptogenesis scenarios,''
  Phys.\ Lett.\ B {\bf 384} (1996) 169
  [hep-ph/9605319].
\bibitem{Flanz:1996fb}
  M.~Flanz, E.~A.~Paschos, U.~Sarkar and J.~Weiss,
  ``Baryogenesis through mixing of heavy Majorana neutrinos,''
  Phys.\ Lett.\ B {\bf 389} (1996) 693
  [hep-ph/9607310].
\bibitem{Enomoto:2013mla}
  S.~Enomoto, S.~Iida, N.~Maekawa and T.~Matsuda,
  ``Beauty is more attractive: particle production and moduli trapping with higher dimensional interaction,''
  JHEP {\bf 1401} (2014) 141
  [arXiv:1310.4751 [hep-ph]].
\bibitem{Enomoto:2014hza}
  S.~Enomoto, N.~Maekawa and T.~Matsuda,
  ``Preheating with higher dimensional interaction,''
  Phys.\ Rev.\ D {\bf 91} (2015) no.10,  103504
  [arXiv:1405.3012 [hep-ph]].
\bibitem{Enomoto:2014cna}
  S.~Enomoto, O.~Fuksinska and Z.~Lalak,
  ``Influence of interactions on particle production induced by time-varying mass terms,''
  JHEP {\bf 1503} (2015) 113
  [arXiv:1412.7442 [hep-ph]].
\bibitem{Sakharov}
 A. D. Sakharov ,"Violation of CP invariance, C asymmetry, and baryon
	asymmetry of the universe",
 Journal of Experimental and Theoretical Physics Letters. 5 (1967) 24-27
\bibitem{Dolgov:1994zq}
  A.~Dolgov and K.~Freese,
  ``Calculation of particle production by Nambu Goldstone bosons with application to inflation reheating and baryogenesis,''
  Phys.\ Rev.\ D {\bf 51} (1995) 2693
  [hep-ph/9410346].
\bibitem{Funakubo:2000us} 
  K.~Funakubo, A.~Kakuto, S.~Otsuki and F.~Toyoda,
  ``Charge generation in the oscillating background,''
  Prog.\ Theor.\ Phys.\  {\bf 105}, 773 (2001)
  [hep-ph/0010266].
\end{thebibliography}
\end{document}